\documentclass[3p,times,procedia]{elsarticle}
\usepackage{nupha_ecrc}

\volume{00}

\firstpage{1}

\journalname{Nuclear Physics A}

\runauth{}

\jid{nupha}

\jnltitlelogo{Nuclear Physics A}

\usepackage{amssymb}
\usepackage{lineno}
\usepackage{amsmath}

 \biboptions{compress}

\usepackage[figuresright]{rotating}

\newcommand{\bfg }{\begin{figure}[htpb]}
\newcommand{\efg }{\end{figure}}
\newcommand{\bmn }{\begin{minipage}}
\newcommand{\emn }{\end{minipage}}
\newcommand{\bt }{\begin{table}[htpb]}
\newcommand{\et }{\end{table}}

\newcommand{\GeVc}{GeV/$c$ }
\newcommand{\GeVcsq}{GeV/$c^2$ }

\newcommand {\pt}   {p_{T}}

\newcommand {\psiPP}    {\Psi_{\rm PP}}
\newcommand {\psiRP}    {\Psi_{\rm RP}}

\newcommand {\psiTPC}    {\Psi_{\rm TPC}}
\newcommand {\psiZDC}    {\Psi_{\rm ZDC}}

\newcommand {\Bvec} {\vec{B}}

\newcommand {\gSS}  {\gamma_{\rm SS}}
\newcommand {\gOS}  {\gamma_{\rm OS}}
\newcommand {\gdel} {\Delta\gamma}
\newcommand {\dg}   {\Delta\gamma}

\newcommand {\note}[1]  {}

\begin{document}

\begin{frontmatter}



\dochead{XXVIIth International Conference on Ultrarelativistic Nucleus-Nucleus Collisions\\ (Quark Matter 2018)}

\title{Measurements of the chiral magnetic effect with background isolation in 200~GeV Au+Au collisions at STAR}

\author{Jie Zhao (for the STAR collaboration)}

\address{Department of Physics and Astronomy, Purdue University, West Lafayette, IN, 47906, USA}

\begin{abstract}

Using two novel methods, pair invariant mass ($m_{inv}$) and comparative measurements with respect to reaction plane ($\Psi_{\rm RP}$) and participant plane ($\Psi_{\rm PP}$), we isolate the possible chiral magnetic effect (CME) from backgrounds in 200 GeV Au+Au collisions at STAR.
The invariant mass method identifies the resonance background contributions, coupled with the elliptic flow ($v_{2}$),
to the charge correlator CME observable ($\Delta\gamma$).
At high mass ($m_{inv}>1.5$~GeV/$c^{2}$) where resonance contribution is small,
we obtain the average $\Delta\gamma$ magnitude.
In the low mass region ($m_{inv}<1.5$~GeV/$c^{2}$),
resonance peaks are observed in $\Delta\gamma$($m_{inv}$). 
An event shape engineering (ESE) method is used to 
model the background shape in $m_{inv}$ to extract the potential CME signal at low $m_{inv}$.
In the comparative method, the $\Psi_{\rm RP}$ is assessed by spectator neutrons measured by ZDC,
and the $\Psi_{\rm PP}$ by the 2$^{nd}$-harmonic event plane measured by the TPC.
The $v_{2}$ is stronger along $\Psi_{\rm PP}$ and weaker along $\Psi_{\rm RP}$;
in contrast, the magnetic field, mainly from spectator protons, is weaker along $\Psi_{\rm PP}$ and stronger along $\Psi_{\rm RP}$.
As a result, the $\Delta\gamma$ measured with respect to $\Psi_{\rm RP}$ and $\Psi_{\rm PP}$ contain different amounts of CME and background,
and can thus determine these two contributions. 
It is found that the possible CME signals with background isolation by these two novel methods are small, on the order of a few percent of the inclusive $\Delta\gamma$ measurements.

\end{abstract}


\end{frontmatter}



\section{Introduction}
Quark interactions with topological gluon fields can induce chirality imbalance and local parity violation 
in quantum chromodynamics (QCD)~\cite{Kharzeev:1998kz}.  
In relativistic heavy-ion collisions, this can lead to observable electric charge separation along the strong magnetic field, $\Bvec$,
produced by spectator protons~\cite{Fukushima:2008xe}.
This is called the chiral magnetic effect (CME).                                                                                              
The commonly used observable to search for the CME-induced charge separation is the three-point azimuthal correlator difference~\cite{Voloshin:2004vk}, $\dg\equiv\gOS-\gSS$
; $\gamma=\langle\cos(\phi_{\alpha}+\phi_{\beta}-2\psiRP)\rangle \approx \langle \cos(\phi_{\alpha} + \phi_{\beta} - 2\phi_{c})\rangle/v_{2}$, 
where $\phi_{\alpha}$ and $\phi_{\beta}$ are the azimuthal angles of two charged particles,
of opposite electric charge sign (OS) or same sign (SS),
and $\psiRP$
is that of the 
reaction plane (span by the impact parameter direction and the beam) to which $\Bvec$ is perpendicular on average.
The latter is often surrogated by the azimuthal angle of a third particle, $\phi_c$, with a resolution correction factor given by the particle's elliptic anisotropy ($v_{2}$).
Significant $\dg$ has indeed been observed in heavy-ion collisions~\cite{Kharzeev:2015znc}.
One of the difficulties in its CME interpretation is a major background contribution
arising from the coupling of
resonance decay correlations and the $v_{2}$ stemming from the participant geometry~\cite{Wang:2009kd,Bzdak:2009fc,Schlichting:2010qia,Adamczyk:2013kcb,Wang:2016iov}.

\section{Invariant mass dependence of the $\dg$ correlator}
\begin{figure}[htbp!]
	\centering 
	\includegraphics[width=4.8cm]{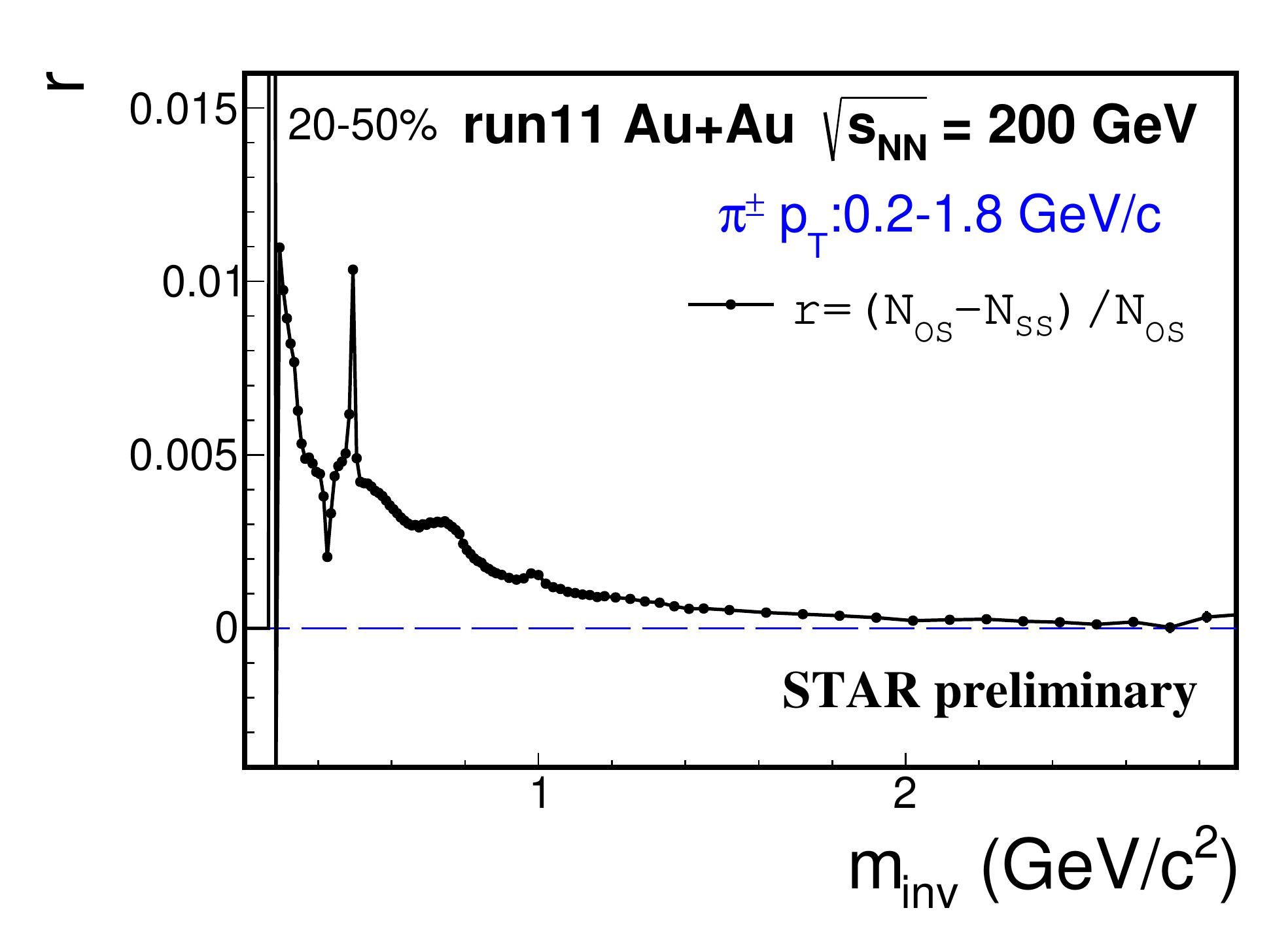} 
	\includegraphics[width=4.8cm]{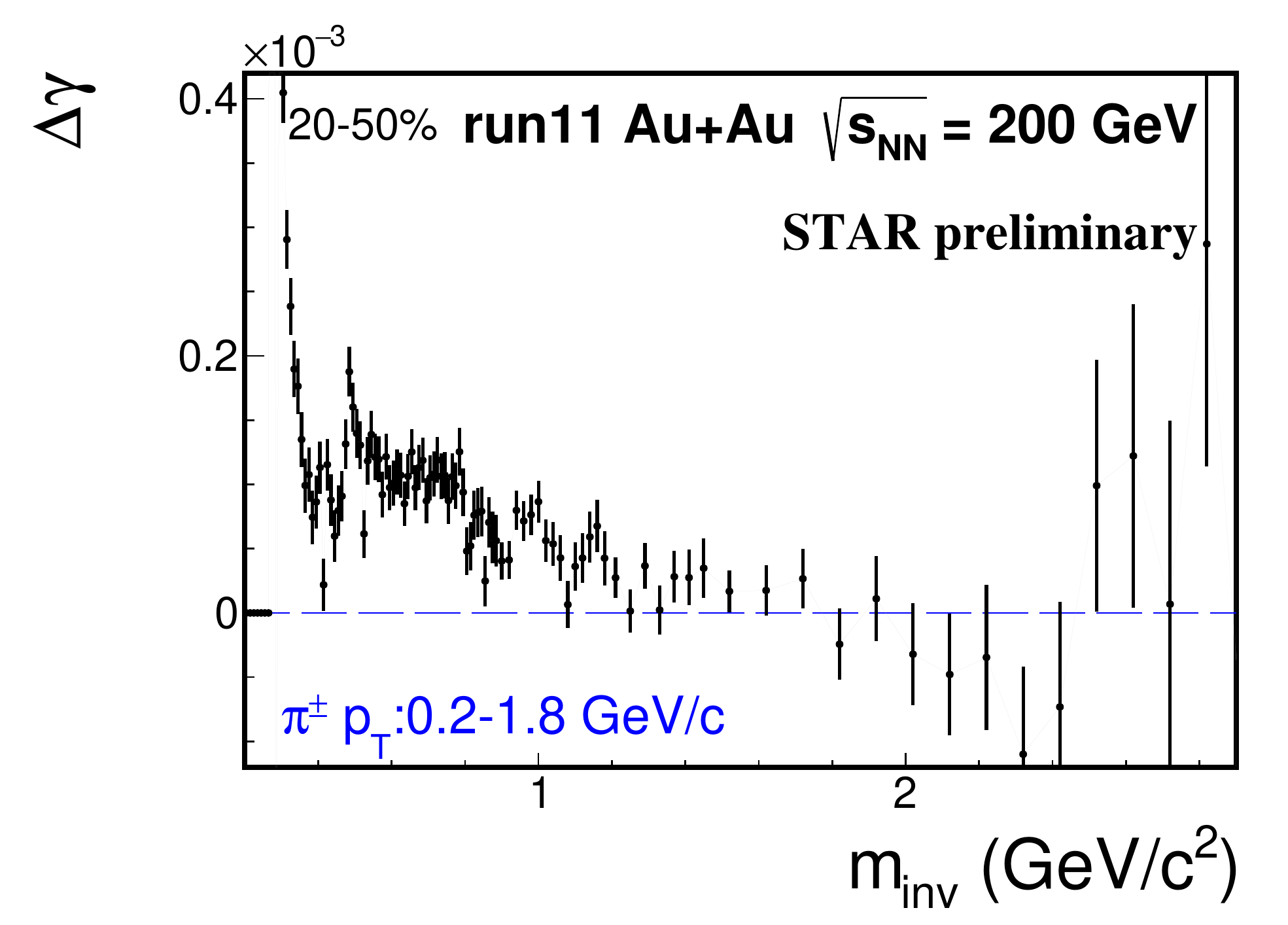} 
	\includegraphics[width=4.8cm]{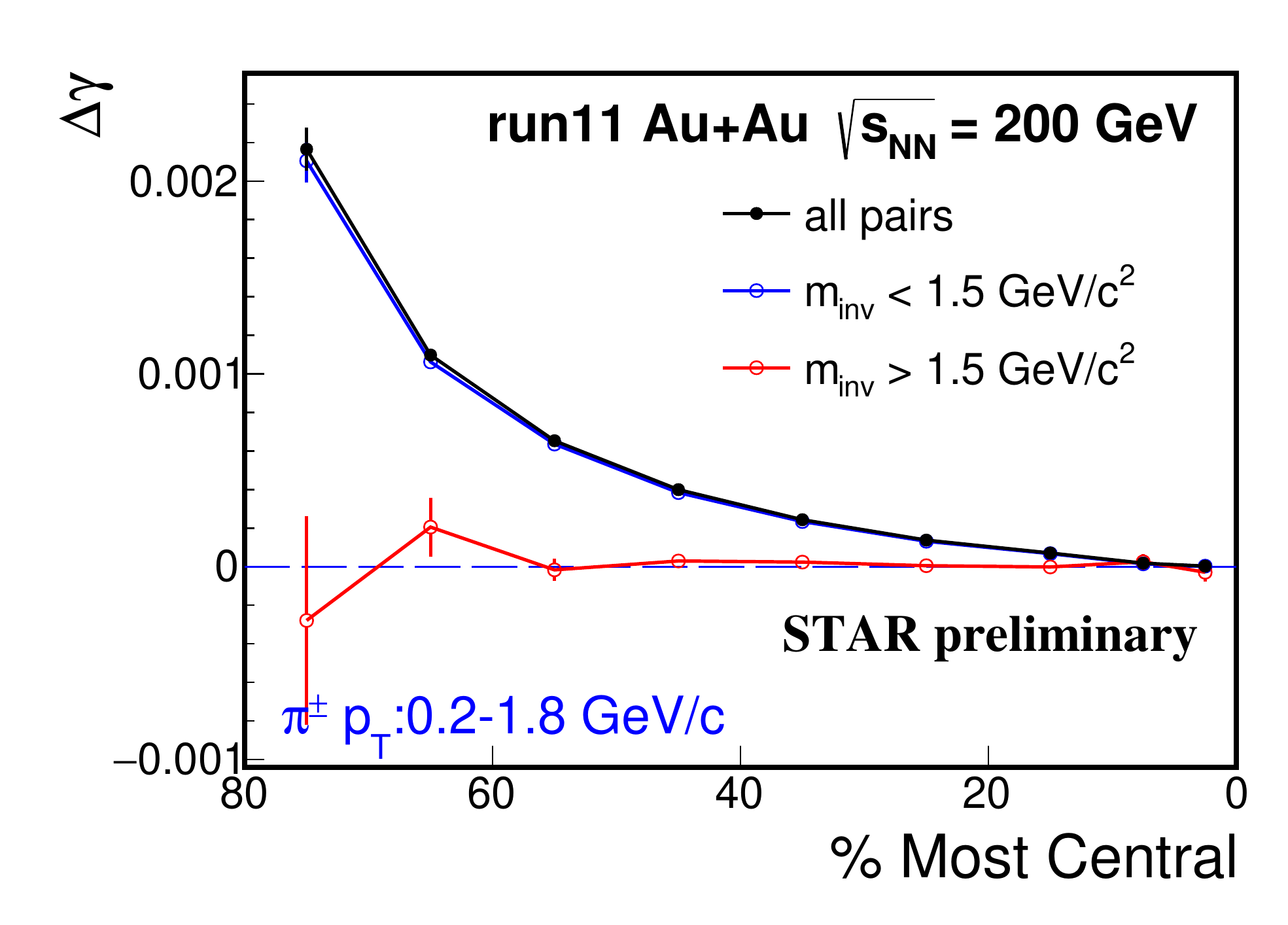} 
	\caption{ 
		Pair invariant mass, $m_{inv}$, dependence of the relative excess of OS over SS pairs, $\rm r=(N_{OS}-N_{SS})/N_{OS}$ (left panel), three-point correlator difference, $\gdel=\gOS-\gSS$ (middle panel). 
		(Right panel) The inclusive $\gdel$ over all mass (black) and at $m_{inv} > 1.5$ \GeVcsq (red) as a function of centrality. The $\pi$ are identified by STAR TPC and TOF with $p_{T}$ from 0.2 to 1.8 \GeVc. 
	}
	\label{CMEmassHigh} 
\end{figure}

The main backgrounds for the $\dg$ are from the resonance decays coupled with $v_{2}$. 
A new analysis approach exploiting the particle pair invariant mass, $m_{inv}$, 
to identify the backgrounds and, hence, to extract the possible CME signal is proposed~\cite{Zhao:2017nfq}. 
Figure~\ref{CMEmassHigh}~(left panel) shows the $m_{inv}$ dependence of the relative excess of OS over SS charged $\pi$ pairs, $\rm r=(N_{OS}-N_{SS})/N_{OS}$,
and (middle panel) shows the $m_{inv}$ dependence of the three-point correlator difference, $\gdel=\gOS-\gSS$.
A lower cut on $m_{inv}$ was used to suppress the resonance contributions.
Figure~\ref{CMEmassHigh}~(right panel) shows the inclusive $\gdel$ over all mass (black) and at $m_{inv} > 1.5$ \GeVcsq (red) as a function of centrality in Au+Au collisions at 200 GeV. 
In 20-50$\%$ collisions centrality, 
combining results from Run-11 ($\sim$0.5 billion minimum-bias events, year~2011), Run-14 ($\sim$0.8 billion, year~2014) and Run-16 ($\sim$1.2 billion, year~2016),
the $\dg$ at $m_{inv} > 1.5$ \GeVcsq is $(5\pm2\pm4)\%$ of the inclusive $\dg$.

The CME is expected to be a low $\pt$ phenomenon~\cite{Kharzeev:2007jp}; its contribution to high mass may be small. 
To extract CME at low mass, resonance contributions need to be subtracted. 
The inclusive $\dg$ can be expressed as $\rm \dg(\it m_{inv}) = r(\it m_{inv})\times \rm cos(\phi_{\alpha} + \phi_{\beta} -2\phi_{reso.})\times v_{2,reso.} + \dg_{\rm CME}$ ~\cite{Zhao:2017nfq}.
The event shape engineering (ESE)~\cite{Schukraft:2012ah} method provides a tool to select events with different $v_{2}$ values by cutting on the $\rm q_{2}$ ($\vec{\rm q}_{2} \rm = 1/N \times \sum(cos(2\phi),sin(2\phi))$).
The difference of the $\dg(m_{inv})$ from different $\rm q_{2}$ classes can be regarded as  
the background $\dg(m_{inv})$ shape~\cite{Zhao:2018ixy}, 
assuming the CME are the same for events from different $\rm q_{2}$ classes.
%

\begin{figure}[htbp!]
	\centering                                                                                                                                    
	\includegraphics[width=4.9cm]{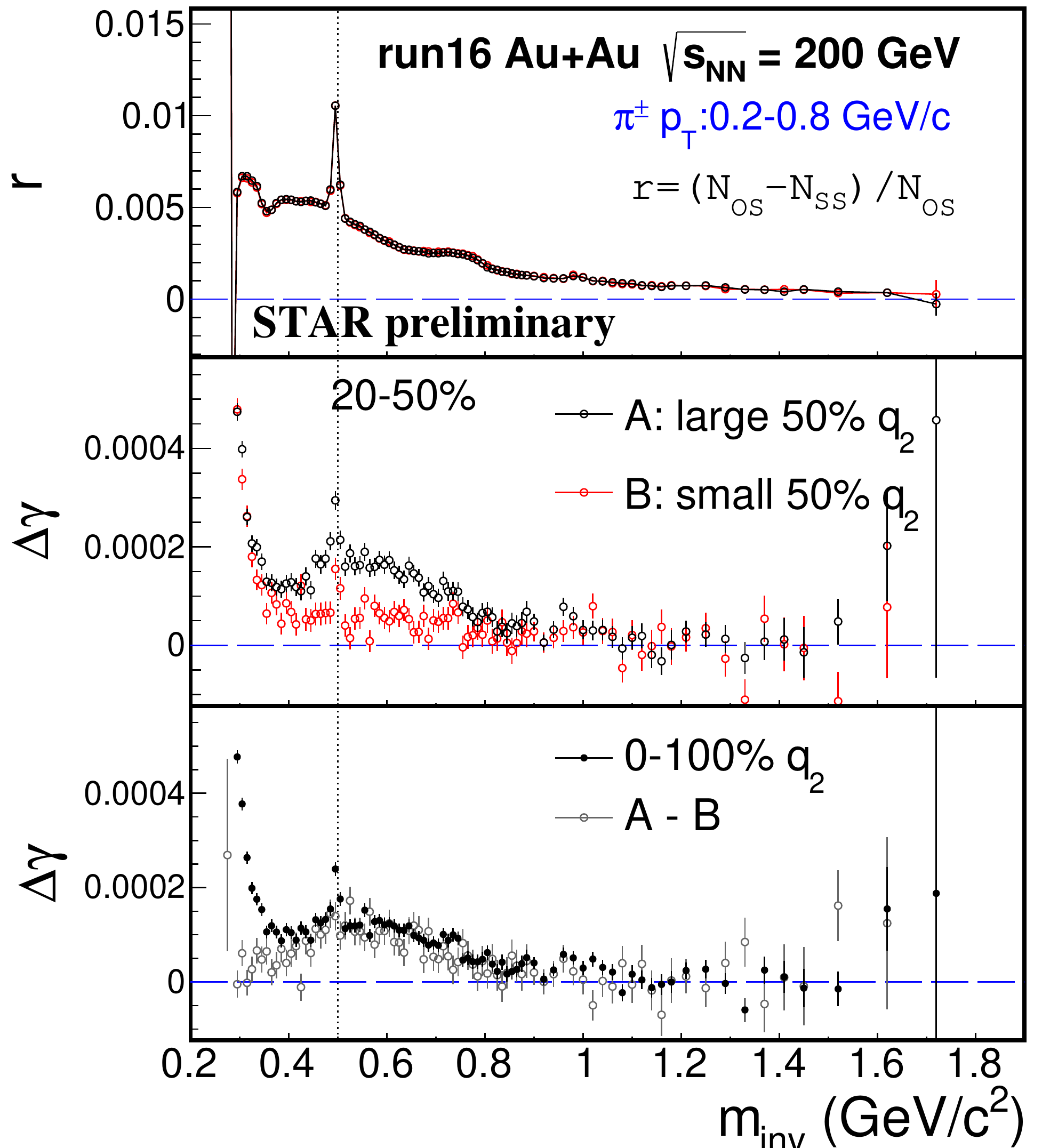}
	\includegraphics[width=5.9cm]{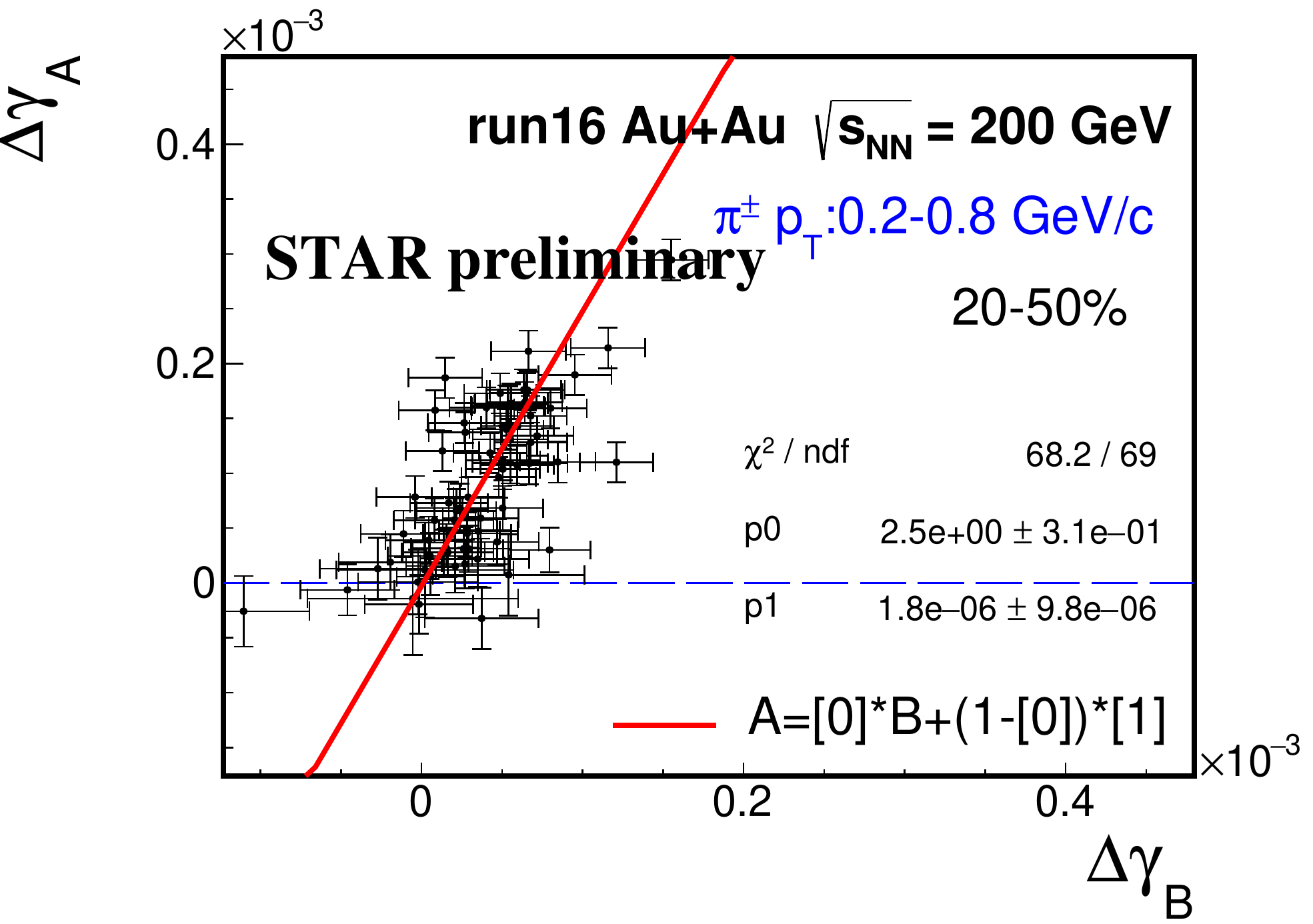}
	\caption{
		Pair invariant mass, $m_{inv}$, dependence of the $\rm r=(N_{OS}-N_{SS})/N_{OS}$ (left top). The $\dg$($m_{inv}$) from ESE selected event sample A (large 50$\%$ $\rm q_{2}$) and B (small 50$\%$ $\rm q_{2}$) (left middle).
		The inclusive (0-100$\%$ $\rm q_{2}$) $\dg$ compared with the difference between $\dg$ from event sample A and B ($\rm \dg_{A} - \dg_{B}$) (left bottom).
		(Right panel) $\rm \dg_{A}$ vs. $\rm \dg_{B}$ fitted by a linear function.
		The $\pi$ are identified by STAR TPC with $p_{T}$ from 0.2 to 0.8 \GeVc.
	}
	\label{CMEmassESE}
\end{figure}

Figure~\ref{CMEmassESE}~(left top) shows the $m_{inv}$ dependence of the $\rm r=(N_{OS}-N_{SS})/N_{OS}$,
(left middle) shows the $\dg$($m_{inv}$) from ESE selected event sample A (large 50$\%$ $\rm q_{2}$) and B (small 50$\%$ $\rm q_{2}$).
The correlators are calculated by $\rm \gamma = cos(\phi_{\alpha} + \phi_{\beta} - 2\phi_{c})/v_{2,c}$. 
The TPC full-event is divided into east and west sub-event, with $\alpha$, $\beta$ and $\rm q_{2}$ from one sub-event and $c$ from other sub-event. 
Figure~\ref{CMEmassESE}~(left bottom) shows the inclusive (0-100$\%$ $\rm q_{2}$) $\dg$ compared with the $\dg$ difference between event samples A and B.
A linear function, $\rm \dg_{A} = b \times \dg_{B} + (1-b) \times \dg_{\rm CME}$, is used to extract the CME.
Figure~\ref{CMEmassESE}~(right) shows the fit result.
Combining Runs 11, 14 and 16, the fit parameter $\dg_{\rm CME}$ is $(2\pm4\pm6)\%$ of inclusive $\dg$ in 20-50$\%$ centrality Au+Au collisions.

\section{$\dg$ with respect to $\Psi_{\rm RP}$ (ZDC) and $\Psi_{\rm PP}$ (TPC)}
The CME-driven charge separation is along the magnetic field direction ($\Psi_{\rm B}$).
The major background to the CME is related to the elliptic flow anisotropy ($v_{2}$), determined by the participant geometry.
A novel idea of differential measurements with respect to the reaction plane ($\psiRP$) and participant plane ($\psiPP$) is proposed~\cite{Xu:2017qfs,Xu:2017zcn}, 
where the $\psiRP$ could be assessed by spectator neutrons measured by the zero-degree calorimeters (ZDC)~\cite{Adler:2001fq}. 
The $v_{2}$ is stronger along $\psiPP$ and weaker along $\psiRP$; in contrast, the magnetic field, being from spectator protons, is weaker along
$\psiPP$ and stronger along $\psiRP$. 
The $\dg$ measured with respect to $\psiRP$ and $\psiPP$ contain different amounts of CME and background, 
and can thus determine these two contributions
assuming the CME is proportional to the magnetic field squared and background is proportional to $v_{2}$
~\cite{Xu:2017qfs}: 
\begin{equation}
	\begin{split}
	& \rm \dg\{\psiTPC\} = \dg_{CME}\{\psiTPC\} + \dg_{Bkg}\{\psiTPC\}, \dg\{\psiZDC\} = \dg_{CME}\{\psiZDC\} + \dg_{Bkg}\{\psiZDC\}, \\
	& \rm \dg_{CME}\{\psiTPC\} = \it{a}*\rm \dg_{CME}\{\psiZDC\}, \dg_{Bkg}\{\psiZDC\}=\it{a}*\rm \dg_{Bkg}\{\psiTPC\},  \\
			  & \it a=\rm v_{2}\{\psiZDC\}/v_{2}\{\psiTPC\}, A=\dg\{\psiZDC\}/\dg\{\psiTPC\},  \\
			  & \rm f^{EP}_{CME} = \dg_{CME}\{\psiTPC\}/\dg\{\psiTPC\} = \it (A/a-1)/(1/a^{2}-1). 
	\end{split}
	\label{eqThreeCtor3}
\end{equation}

\begin{figure}[htbp!]
	\centering
	\includegraphics[width=4.8cm]{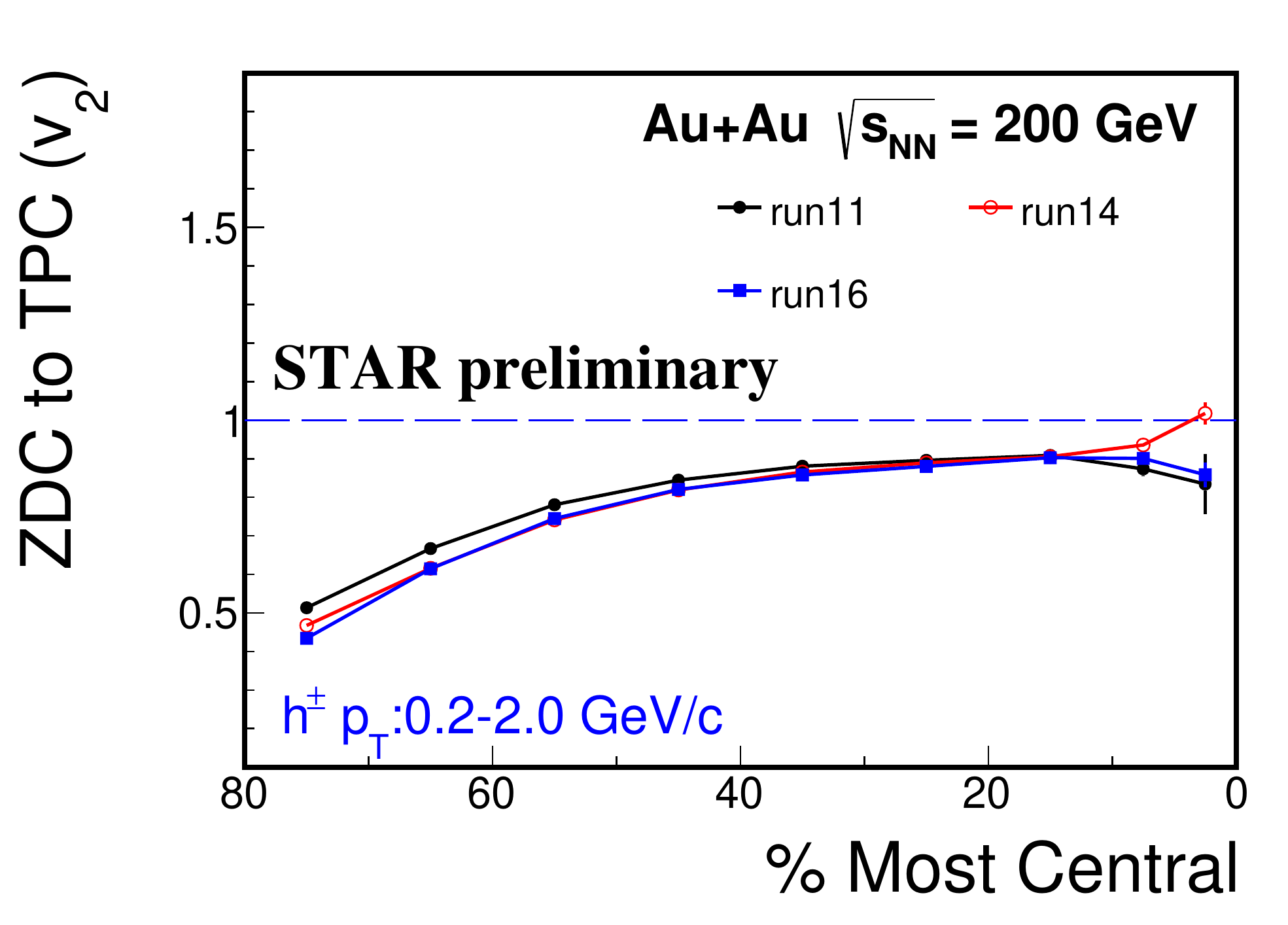}
	\includegraphics[width=4.8cm]{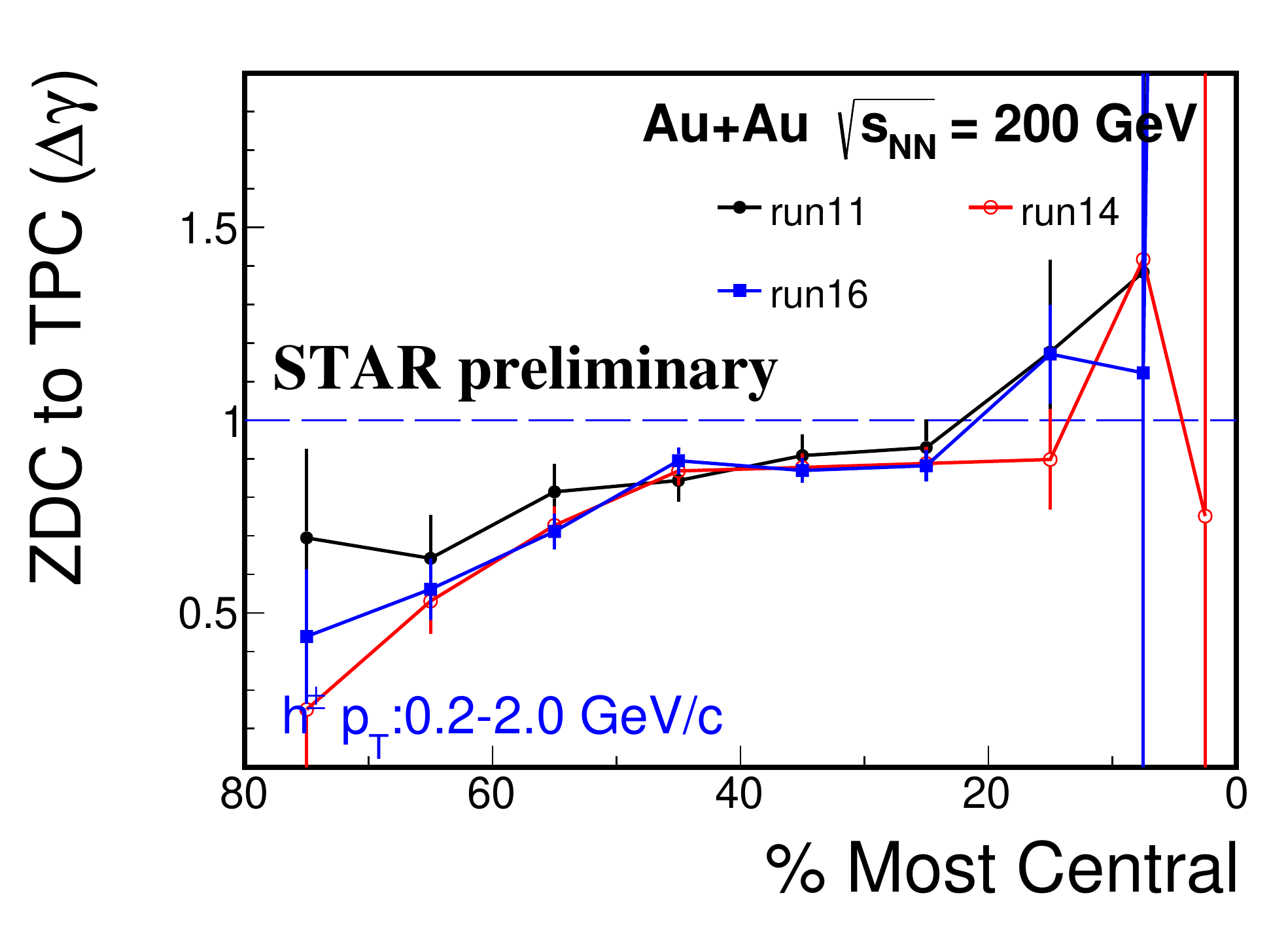}
	\includegraphics[width=4.8cm]{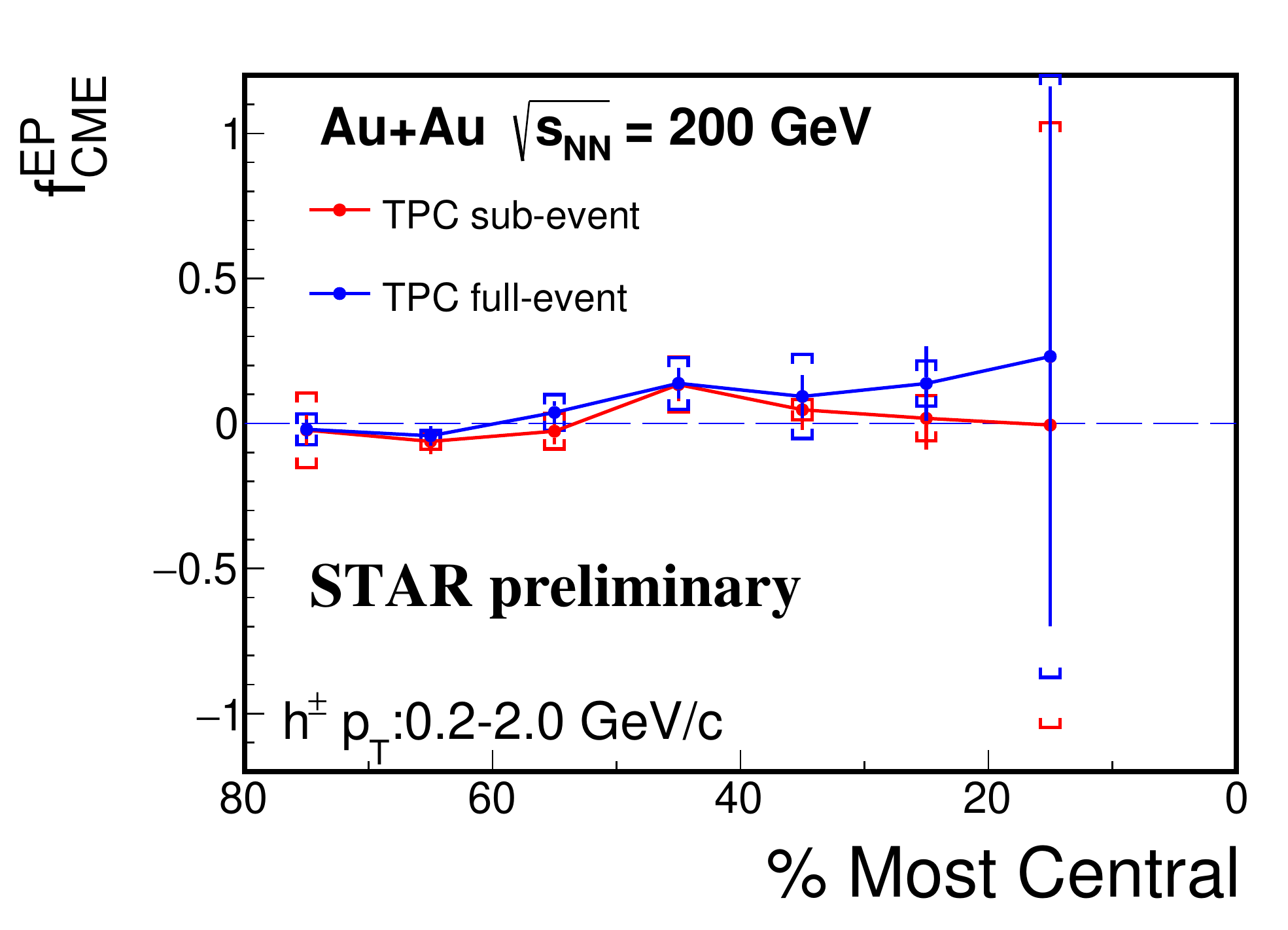}
	\caption{
		The centrality dependences of the ratios of the $v_{2}$ (left panel) and $\dg$ (middle panel) 
		measured with respect to the ZDC event plane to those with respect to the TPC event plane.
		(Right panel) The extracted $\rm f^{EP}_{CME}$ as a function of collision centrality.
	}
	\label{CMEZDC}
\end{figure}

Figure~\ref{CMEZDC} shows the ratio of $v_{2}$ (left)
measured with respect to the ZDC event plane and the $v_{2}$ with respect to the TPC event plane, $a=\rm v_{2}\{\psiZDC\}/v_{2}\{\psiTPC\}$ in Eq.~\ref{eqThreeCtor3}, 
and that of $\dg$ (middle panel), $A=\dg\{\psiZDC\}/\dg\{\psiTPC\}$ in Eq.~\ref{eqThreeCtor3}, 
as functions of collisions centrality. To suppress the non-flow contributions in $v_{2}$ and $\dg$ measurements, 
the TPC sub-event method is used, where each TPC event is divided into east and west sub-event, 
with event-plane from one sub-event and particles of interest from the other sub-event. 
Figure~\ref{CMEZDC}~(right) show the extracted possible CME fraction ($\rm f^{EP}_{CME}$)~\cite{Xu:2017qfs} as function of centrality. 
For comparison the results from TPC full-event method are also plotted.
The extracted $\rm f^{EP}_{CME}$ (combined from Runs 11, 14 and 16) are $(9\pm4\pm7)\%$ and $(12\pm4\pm11)\%$ 
from the TPC sub-event and full-event methods in 20-50$\%$ centrality Au+Au collisions, respectively.

\section{Summary}
Charge separation measurements by the three-point azimuthal correlator ($\dg$) are contaminated by major backgrounds arising from resonance decay 
correlations coupled with the elliptical anisotropy ($v_2$). 
To reduce/eliminate background contaminations, 
two novel methods are employed:
the $\dg$ correlator as a function of the particle pair invariant mass ($m_{inv}$) and 
the comparative $\dg$ measurements with respect to $\Psi_{\rm RP}$ (estimated by ZDC) and $\Psi_{\rm PP}$ (estimated by TPC).

Resonance structures are observed in $\dg$ as function of $\pi$-$\pi$ $m_{inv}$.
A lower $m_{inv}$ cut ($m_{inv}>1.5$~\GeVcsq) yields a $\dg$ fraction of $(5\pm2\pm4)\%$ of the inclusive $\dg$ measurement. 
In the low mass region, event shape engineering is used to determine the background shape in $m_{inv}$, 
and a linear fit to $\dg$($m_{inv}$) yields a possible CME signal of $(2\pm4\pm6)\%$ of the inclusive $\dg$ measurement in 20-50$\%$ centrality Au+Au collisions.

The $\dg$ measurements with respect to $\Psi_{\rm RP}$ and $\Psi_{\rm PP}$ contain different amounts of CME and background. 
The $v_{2}$ is stronger along $\psiPP$ and weaker along $\psiRP$; and the magnetic field is weaker along
$\psiPP$ and stronger along $\psiRP$.
By comparing the $v_{2}$ and $\dg$ with respect to $\Psi_{\rm RP}$ and $\Psi_{\rm PP}$, the extracted possible CME fractions
are $(9\pm4\pm7)\%$ and $(12\pm4\pm11)\%$ from the TPC sub-event and full-event methods in 20-50$\%$ centrality Au+Au collisions, respectively.

\begin{figure}[htbp!]
	\centering 
	\includegraphics[width=7.3cm]{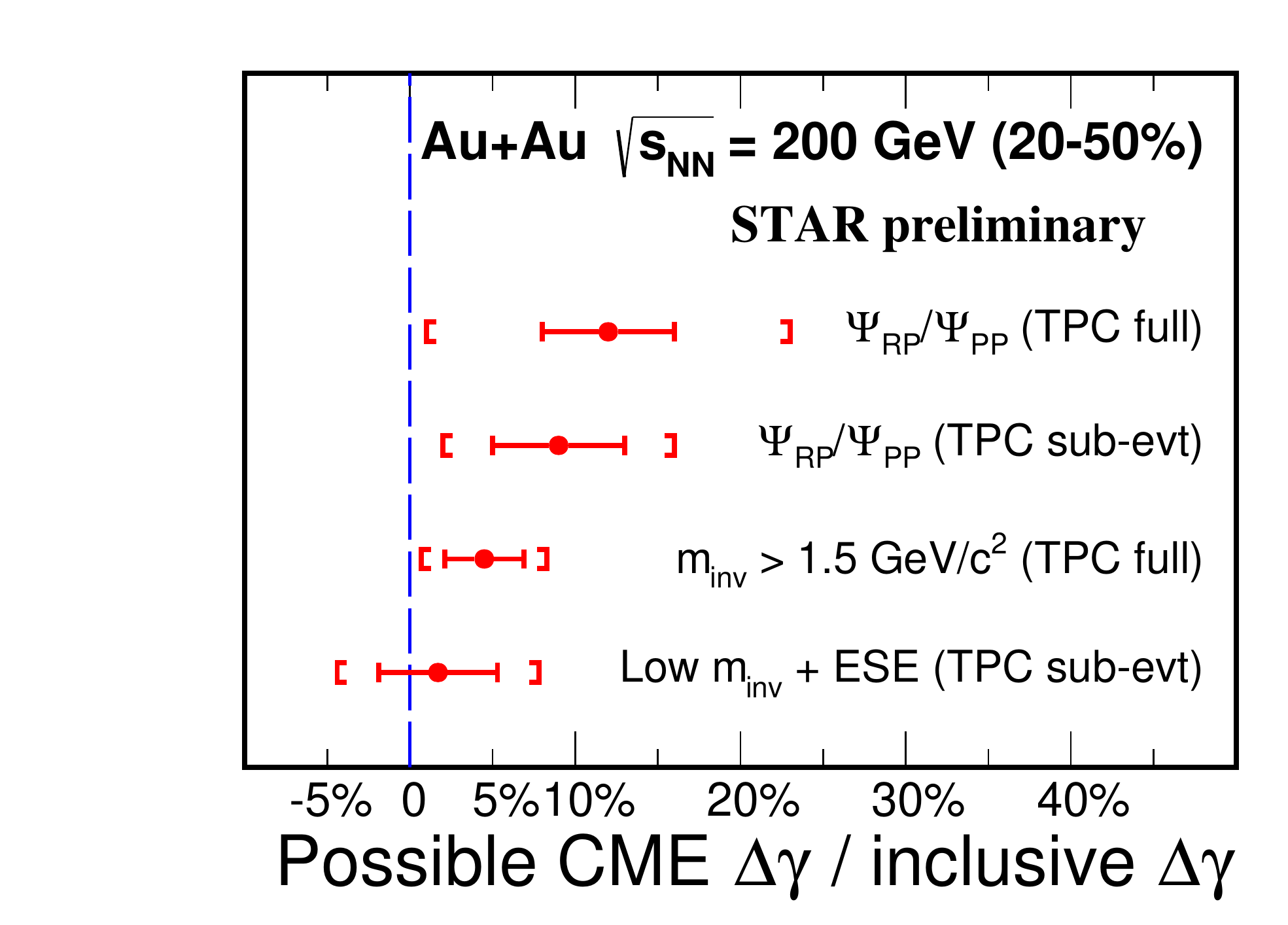} 
	\caption{
		The possible CME $\dg$ over the inclusive $\dg$ fraction from different analysis methods in middle central (20-50$\%$) Au+Au collisions at $\sqrt{s_{NN}}$ = 200 GeV.
	}
	\label{CMEsummary} 
\end{figure}

The extracted potential CME signal fraction 
(CME $\dg$ over the inclusive $\dg$) 
in middle central (20-50$\%$) Au+Au collisions at $\sqrt{s_{NN}}$ = 200 GeV are summarized in Fig.~\ref{CMEsummary}.
These data-driven estimates indicate that the possible CME signal is small, within 1-2 $\sigma$ from zero.
Precision can be improved in the future with more Au+Au data and the new isobar data. 
Possible ZDC upgrades to achieve better ΨRP determination are being investigated.

\vspace{3 mm}
$\bold{Acknowledgments}$ This work was partly supported by the U.S. Department of Energy (Grant No. de-sc0012910),
and the NSFC of China under Grant No. 11505073.





\bibliographystyle{elsarticle-num}
\bibliography{ref}

\begin{thebibliography}{10}
\expandafter\ifx\csname url\endcsname\relax
  \def\url#1{\texttt{#1}}\fi
\expandafter\ifx\csname urlprefix\endcsname\relax\def\urlprefix{URL }\fi
\expandafter\ifx\csname href\endcsname\relax
  \def\href#1#2{#2} \def\path#1{#1}\fi

\bibitem{Kharzeev:1998kz}
D.~Kharzeev, R.~D. Pisarski, M.~H.~G. Tytgat, Phys. Rev. Lett. 81 (1998)
  512--515.

\bibitem{Fukushima:2008xe}
K.~Fukushima, D.~E. Kharzeev, H.~J. Warringa, Phys. Rev. D78 (2008) 074033.

\bibitem{Voloshin:2004vk}
S.~A. Voloshin, Phys. Rev. C70 (2004) 057901.

\bibitem{Kharzeev:2015znc}
D.~E. Kharzeev, J.~Liao, S.~A. Voloshin, G.~Wang, Prog. Part. Nucl. Phys. 88
  (2016) 1--28.

\bibitem{Wang:2009kd}
F.~Wang, Phys.Rev. C81 (2010) 064902.

\bibitem{Bzdak:2009fc}
A.~Bzdak, V.~Koch, J.~Liao, Phys.Rev. C81 (2010) 031901.

\bibitem{Schlichting:2010qia}
S.~Schlichting, S.~Pratt, Phys.Rev. C83 (2011) 014913.

\bibitem{Adamczyk:2013kcb}
L.~Adamczyk, et~al., Phys. Rev. C89~(4) (2014) 044908.

\bibitem{Wang:2016iov}
F.~Wang, J.~Zhao, Phys. Rev. C95~(5) (2017) 051901 (R).

\bibitem{Zhao:2017nfq}
J.~Zhao, H.~Li, F.~Wang, {}\href {http://arxiv.org/abs/1705.05410}
  {\path{arXiv:1705.05410}}.

\bibitem{Kharzeev:2007jp}
D.~E. Kharzeev, L.~D. McLerran, H.~J. Warringa, Nucl. Phys. A803 (2008)
  227--253.

\bibitem{Schukraft:2012ah}
J.~Schukraft, A.~Timmins, S.~A. Voloshin, Phys. Lett. B719 (2013) 394--398.

\bibitem{Zhao:2018ixy}
J.~Zhao, Int. J. Mod. Phys. A33~(13) (2018) 1830010.

\bibitem{Xu:2017qfs}
H.~Xu, J.~Zhao, X.~Wang, H.~Li, Z.~Lin, C.~Shen, F.~Wang, Chin. Phys. C 42
  (2018) 084103.

\bibitem{Xu:2017zcn}
H.~Xu, X.~Wang, H.~Li, J.~Zhao, Z.~Lin, C.~Shen, F.~Wang, Phys. Rev. Lett. 121
  (2018) 022301.

\bibitem{Adler:2001fq}
C.~Adler, H.~Strobele, A.~Denisov, E.~Garcia, M.~Murray, S.~White, Nucl.
  Instrum. Meth. A461 (2001) 337--340.

\end{thebibliography}








\end{document}